\documentclass[namedreferences]{solarphysics}
\usepackage[optionalrh]{spr-sola-addons} 
\usepackage{graphicx}        
\usepackage{color}           
\usepackage{breakurl}        
\usepackage{hyperref}

\newcommand{\etal}{{\it et al.}}

\newcommand{\apj}{    {\it Astrophys. J.}}
\newcommand{\apjl}{   {\it Astrophys. J. Lett.}}
\newcommand{\apss}{   {\it Astrophys. Space Sci.}}

\newcommand{\jgr}{    {\it J. Geophys. Res.}}
\newcommand{\mnras}{  {\it Mon. Not. Roy. Astron. Soc.}}

\newcommand{\solphys}{{\it Solar Phys.}}
 
\newcommand{\ssr}{    {\it Space Sci. Rev.}} 
\chardef\us=`\_

\begin{document}

\begin{article}
\begin{opening}

\title{Comparing SSN Index to X-ray Flare and Coronal Mass Ejection Rates from Solar Cycles
22\,--\,24}

\author[addressref={aff1},corref,email={lwinter@aer.com}]{\inits{L.M.}\fnm{L.M.}~\lnm{Winter}}
\author[addressref={aff1},corref]{\inits{R.L.}\fnm{R.L.}~\lnm{Pernak}}
\author[addressref={aff2}]{\inits{K.S.}\fnm{K.S.}~\lnm{Balasubramaniam}}


\address[id=aff1]{Atmospheric and Environmental Research, 131 Hartwell Avenue, Lexington, MA 02421, USA}
\address[id=aff2]{Battlespace Environment Division, Kirtland AFB, NM 87117, USA}

\runningauthor{L.M. Winter \etal}
\runningtitle{Sunspot Number and X-ray Flux}

\begin{abstract}
The newly revised sunspot number series allows for placing historical geoeffective storms in the context of several hundred years of solar activity. Using statistical analyses of the {\it Geostationary Operational Environmental Satellites} (GOES) X-ray observations from the past $\approx$\,30 years and the {\it Solar and Heliospheric Observatory} (SOHO) {\it Large Angle and Spectrometric Coronagraph} (LASCO) Coronal Mass Ejection (CME) catalog (1996\,--\,present), we present sunspot-number-dependent flare and CME rates. In particular, we present X-ray flare rates as a function of sunspot number for the past three cycles.  We also show that the 1\,--\,8~\AA~X-ray background flux is strongly correlated with sunspot number across solar cycles. Similarly, we show that the CME properties ({\it e.g.}, {proxies related to the CME} linear speed and width) are also correlated with sunspot number for SC 23 and 24. These updated rates will enable future predictions for geoeffective events and place historical storms in the context of present solar activity. 
\end{abstract}
\end{opening}

\section{Introduction}
     \label{S-Introduction} 

\indent\indent  Cyclical variability of the Sun, first discovered through analyses of the sunspot number \citep{1852MNRAS..13...29W}, affects the interplanetary/planetary magnetic fields and environment in a number of ways. During solar maximum, increases in flares and coronal mass ejections (CMEs) lead to an increase in the injection, acceleration, and transport of solar energetic particles. Earth-directed solar particle events last from a few minutes to weeks, occurring from a few times a year to once a month. They can damage satellite electronics and even pose health risks to astronauts and aviators
 \citep{2000JGR...10510543F}.
Associated geomagnetic effects from enhanced solar activity disrupt ionospheric communications and can result in high-magnitude geomagnetically induced currents, causing power outages like the March 1989 Quebec Blackout [Details are available in the North American Electric Reliability Council report at \url{http://www.nerc.com/files/1989-Quebec-Disturbance.pdf}.].  During solar minimum, the total heliospheric magnetic flux weakens allowing for an enhancement in the levels of galactic cosmic rays (GCRs), relativistic particles of energies from 100\,MeV through 10\,GeV that are mainly produced in powerful supernova explosions from stellar nurseries throughout the galaxy. Prolonged solar minima, like the recently observed extended minimum between SC 23 and 24, lead to high level radiation dose rates in interplanetary space that are a serious health hazard for astronauts during extended space travel ({\it e.g.}, \citealt{SWE:SWE386}).

In this article, we present an analysis of the X-ray and CME properties spanning several solar cycles. In the X-rays, the {\it Geostationary Operational Environmental Satellites} (GOES) spacecraft provide continuous soft X-ray monitoring data for nearly 40 years. We utilize these data to relate changes in the non-flare solar X-ray background to the sunspot number for SC 22\,--\,24 in Section~\ref{S-background}. In Section~\ref{S-flares}, we expand upon earlier research into variation of X-ray flare rates with solar activity ({\it e.g.}, \citealt{2011SoPh..274...99A}) by characterizing flare rates as a function of sunspot number. In Section~\ref{S-CME} we compare {\it Large Angle and Spectrometric Coronagraph} (LASCO) { measurements related to CME} speed and width to sunspot number for SC 23 and 24. We also compare the CME and X-ray flare properties over these cycles. The results of our analysis include updated flare and CME frequencies that include SC 24 activity. With the newly updated sunspot number index, these new rates can be compared to historical rates to place current solar activity in the context of the historical record. We discuss this further in Section~\ref{S-Conclusion}.

\section{X-ray Background and the Revised SSN}
\label{S-background}

\indent\indent The solar soft X-ray background flux, produced in the corona from free\,--\,free bremsstrahlung and line emission from flare plasma in active region magnetic loops, varies cyclically ({\it e.g.}, \citealt{1988AdSpR...8...67W}, \citealt{1994SoPh..152...53A}, and \citealt{2041-8205-793-2-L45}). In our previous work \citep{2041-8205-793-2-L45}, we utilized the publicly available 1\,--\,8\,\AA~X-ray observations from GOES from 1986\,--\,2014 to establish that this background flux can be used to predict the date of solar cycle maximum, the maximum X-ray background at maximum and throughout the cycle, and the date of the next solar minimum from a few years into the solar cycle. The background is measured as the minimum 1\,--\,8\,\AA~flux in the preceding 24 hours for each one-minute XRS measurement, following the procedure of \citet{SWE:SWE20042}. This technique excludes strong flares and instead measures the quiescent flux from active regions. 

\begin{figure}
\centering
\includegraphics[width=0.9\linewidth]{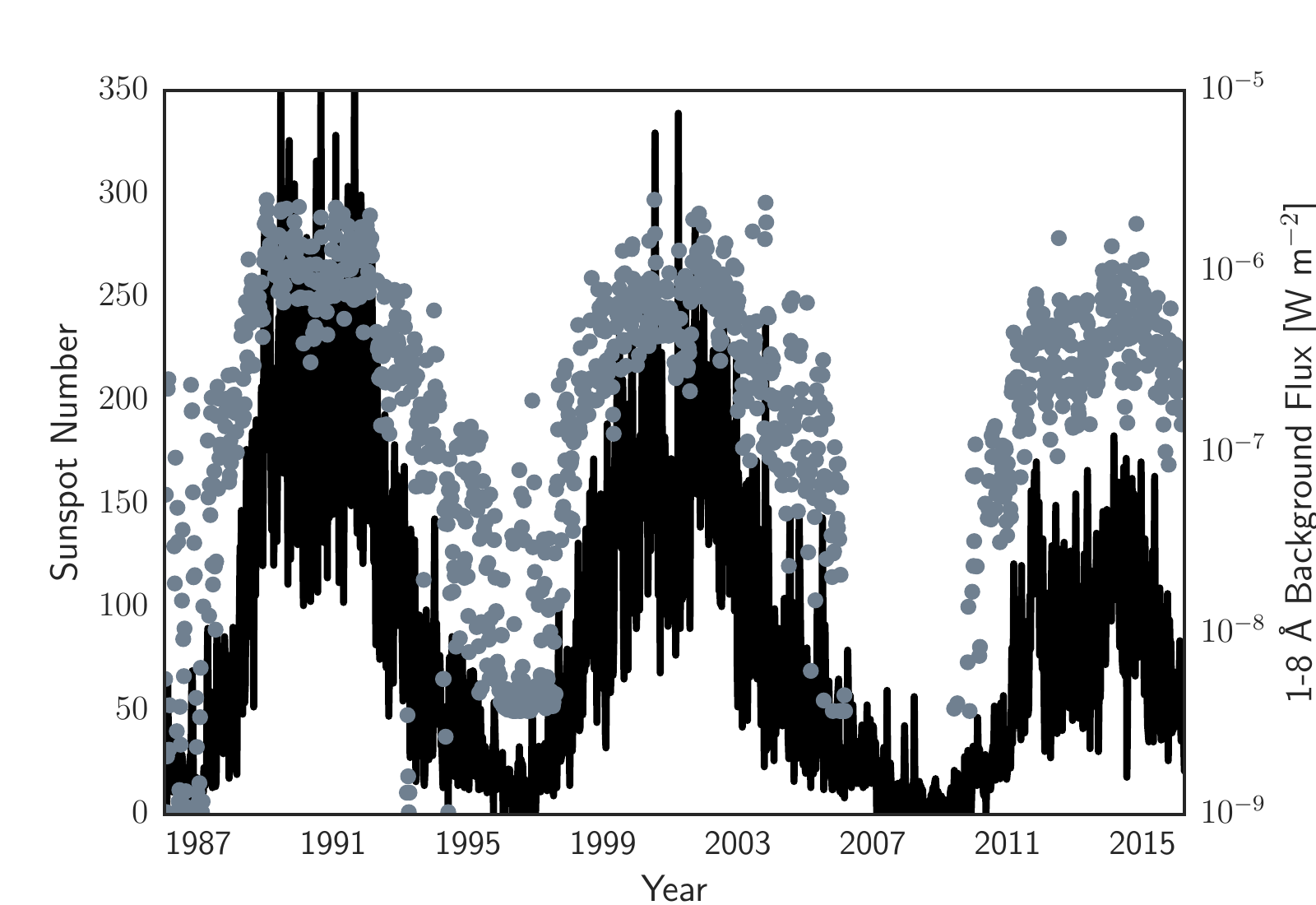}
\caption{Comparison of the revised sunspot number (black) and the 1\,--\,8\,\AA~X-ray background flux (gray) for SC 22\,--\,24. The one-minute X-ray background values are median combined in one-week bin sizes. During the last solar minimum, the background was often below the GOES detection limit. For the sunspot number, the daily total SSN is median combined in one-week bins. The cyclical activity is seen in both the X-ray background and sunspot number, on a similar time scale (as discussed {\it e.g.}, by
\citealt{2041-8205-793-2-L45}).
}
\label{fig-xraybackground}
\end{figure}

Extending upon our previous soft X-ray background study, we compare the 1\,--\,8\,\AA~background flux described by \citet{2041-8205-793-2-L45} with the revised sunspot number \citep{2014SSRv..186...35C}. We included analyses of the GOES 1\,--\,8\,\AA~data from January 1986 through February 2016. The revised sunspot number series was obtained for this same period from the WDC-SILSO, Royal Observatory of Belgium, Brussels obtained from \url{http://www.sidc.be/silso/datafiles}. Figure~\ref{fig-xraybackground} shows the comparison of the revised sunspot number (the daily total sunspot number, binned to a one-week interval) to the X-ray background (binned to a one-week interval). Both parameters characterize the solar activity level throughout the cycles by tracking changes in the active regions ({\it e.g.}, through relative X-ray brightness or number of sunspots).

\begin{figure}
\centering
\includegraphics[width=0.9\linewidth]{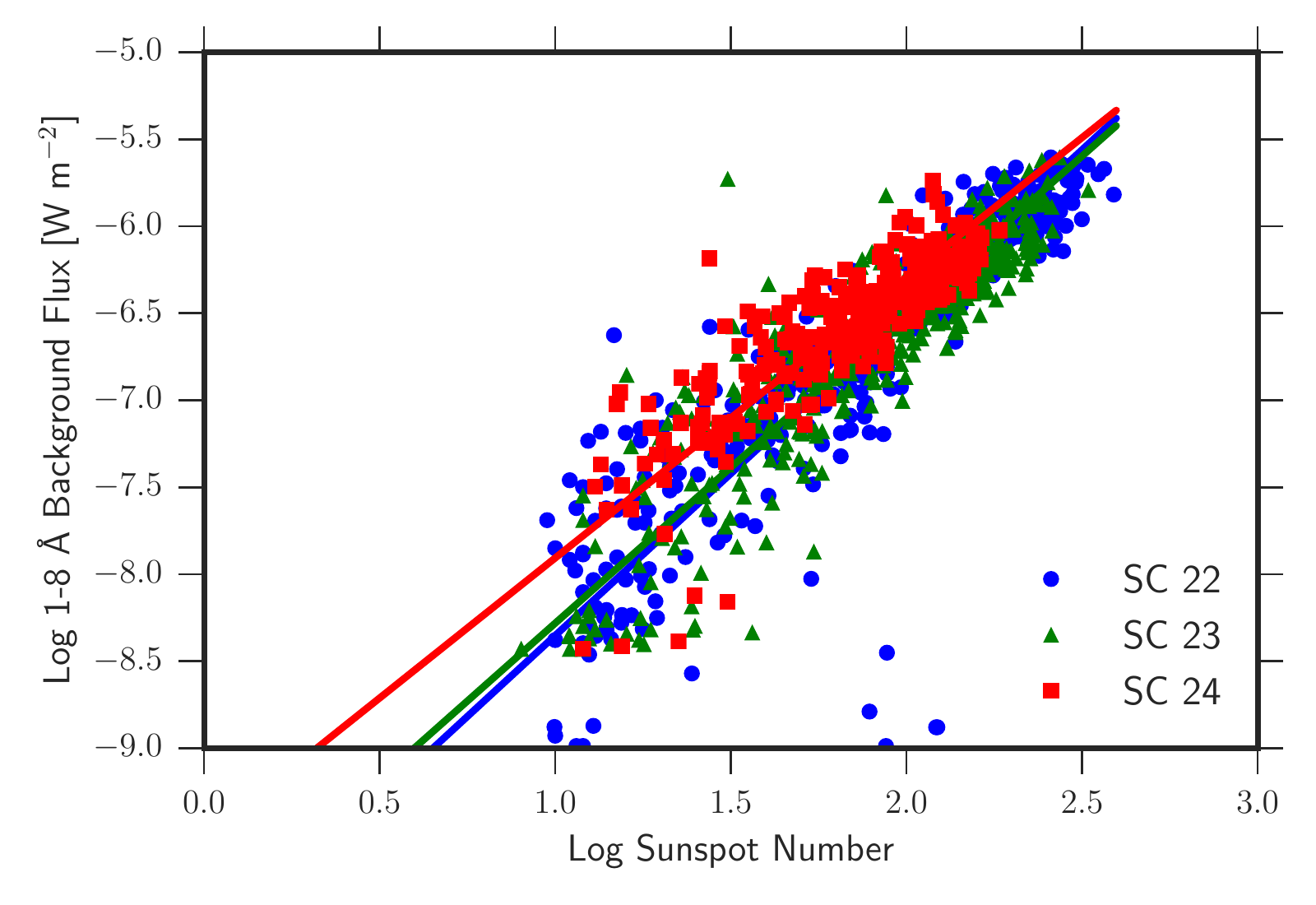}
\caption{Relationship between the revised sunspot number and the 1\,--\,8\,\AA~X-ray flux for SC 22\,--\,24. Both data sets are binned by one-week time intervals. The resulting best-fit linear regression models are included in Table~\ref{table-SSNfits}. The slopes from each of the solar cycles are consistent within the errorbars, with differences in this relationship tied to the normalization factor.
}
\label{fig-ssnxraybackground}
\end{figure}

In Figure~\ref{fig-ssnxraybackground}, we show the direct correlation between X-ray background flux and sunspot number for SC 22\,--\,24. For each solar cycle, we perform a linear regression fit of the form:  ${\log}$\,B = Slope ${\log}$\,SSN + Intercept. Both the background and SSN are binned by one week, with the same dates used for the binning time periods. The best-fit values and associated correlation coefficients are included in Table~\ref{table-SSNfits}. We also computed Spearman rank correlation coefficients for the direct relationship (non-logarithmic) between these values, resulting in $R^2$ of 0.79 (SC 22), 0.65 (SC 23), and 0.59 (SC 24). In both cases, there is a significant correlation between X-ray background and SSN. 

{ Both the slope and the intercept values change between the solar cycles, although we note that these changes are not statistically significant when accounting for the error bars. However, we find that the slope flattens and the intercept increases with decreasing solar cycle strength. This effect is best seen in Figure~\ref{fig-ssnxraybackground} at low sunspot number ($< 30$), where the average X-ray background is higher in the weaker solar cycles (i.e., SC 24) in the stronger solar cycles. Since the background is below the GOES detection limit during the extremely weak minimum between SC 23 and 24, we tested whether this result holds for higher sunspot number where our sampling of X-ray background flux is complete. We find slopes of $2.04 \pm 0.12$ (SC 22), $1.72 \pm 0.06$ (SC 23), and $1.51 \pm 0.07$ (SC 24), with $R^2 \sim 0.6$, where SSN $> 30$. Similarly, we find slopes of 
$2.37 \pm 0.18$ (SC 22), $1.75 \pm 0.06$ (SC 23), and $1.65 \pm 0.09$ (SC 24), with $R^2 > 0.5$, where SSN $> 50$. Therefore, the flatter slope in the relationship between SSN and X-ray background flux for lower activity levels is not an effect of poor measurements of X-ray flux at low SSN.

A possible explanation for the differences in the relation between SSN and X-ray background between solar cycles is that the Sun is further into the activity cycle when lower sunspot numbers are observed during a weaker cycle when compared to the earlier phase of a stronger solar cycle ({\it e.g.}, SSN of 30 is reached closer in time to solar maximum in SC 24 than in either SC 22 or 23). The overall magnetic activity may be higher in the weaker solar cycle, as indicated by a higher X-ray background flux, despite the SSN being the same as during the stronger solar cycle. A study of the size and morphology, as well as estimates of the total energetics, of the sunspots during similar SSN between weaker and stronger solar cycles might reveal further differences.}

These relationships between SSN and X-ray background flux give us a means to estimate the X-ray flux during historical time periods where X-ray measurements were not taken. In turn, in the following section the X-ray activity level is compared to the X-ray flare rate, for which our X-ray background\,--\,SSN relationship allows a comparison of the expected rate for historical time periods. The X-ray background\,--\,SSN relationship can be used as a method of predicting the X-ray flare and CME rates using the SSN from earlier solar cycles where X-ray measurements are not available.

\begin{table}
\caption{Results from the Comparison of the X-ray Background Flux to SSN}\label{table-SSNfits}
\begin{tabular}{lllll}
\hline
{\bf Solar Cycle} & {\bf Slope} & {\bf Intercept} & {\bf R$^2$} & {$\tau$} \\
\hline
22 & 1.86 $\pm$ 0.06 & -10.21 $\pm$ 0.11 & 0.77 & 0.77\\
23 & 1.77 $\pm$ 0.05 & -10.01 $\pm$ 0.10 & 0.80 & 0.72\\
24 & 1.61 $\pm$ 0.09 & -9.52 $\pm$ 0.17 & 0.70 & 0.66\\
\hline
\end{tabular}
\vspace{0.25cm}

{\small
Results from a linear regression fit of the form ${\log}$\,B = Slope ${\log}$\,SSN + Intercept. Error bars are included along with the square of the Spearman's rank and the Kendall's $\tau$ correlation coefficients, which indicate a strong correlation between these parameters. The associated p-value for Kendall's $\tau$ is $\ll 10^{-10}$ in all solar-cycle comparisons.
}
\end{table}


\section{X-ray Flare Rates} 
  \label{S-flares}

\begin{figure}
\centering
\includegraphics[width=\linewidth]{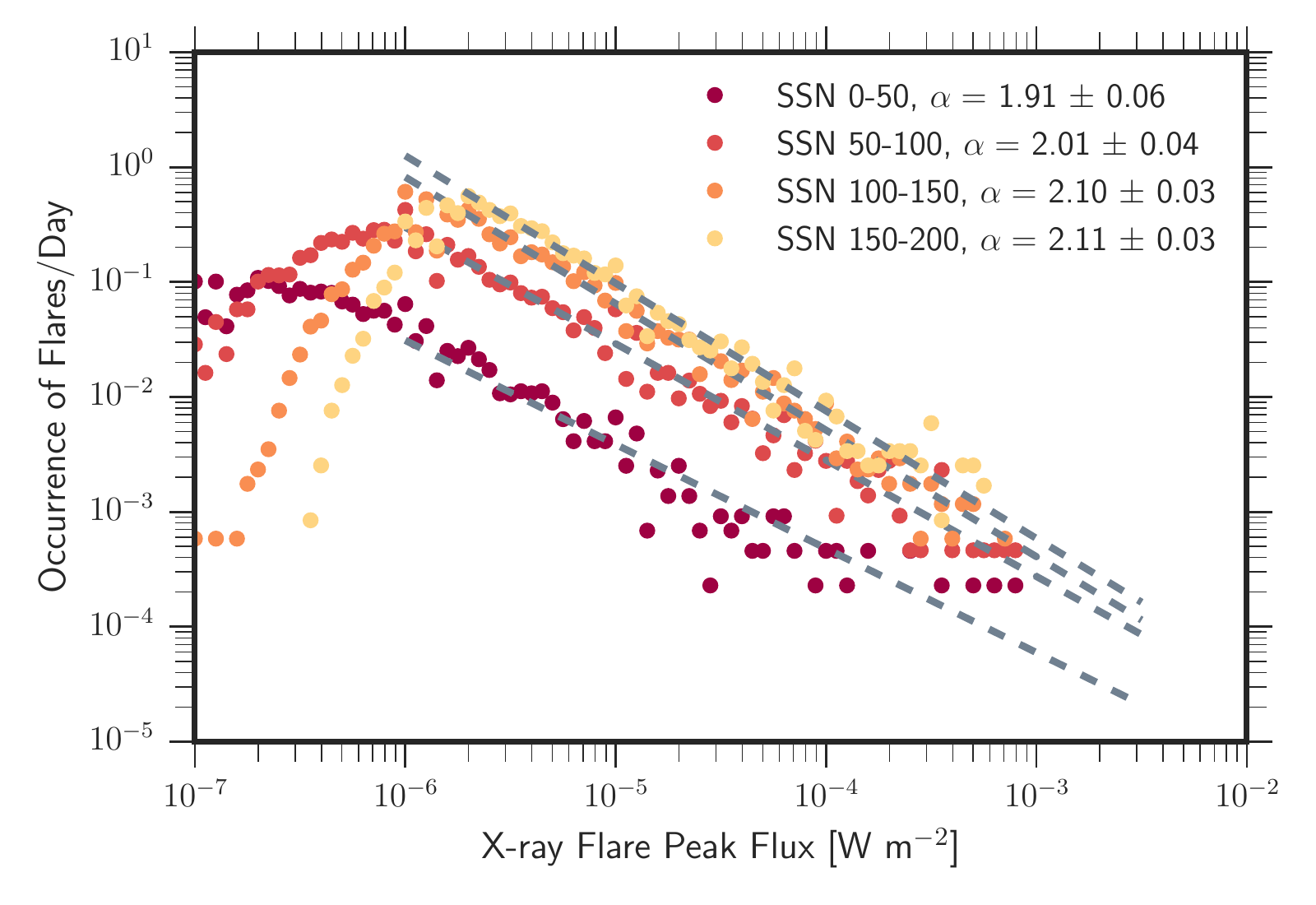}
\includegraphics[width=\linewidth]{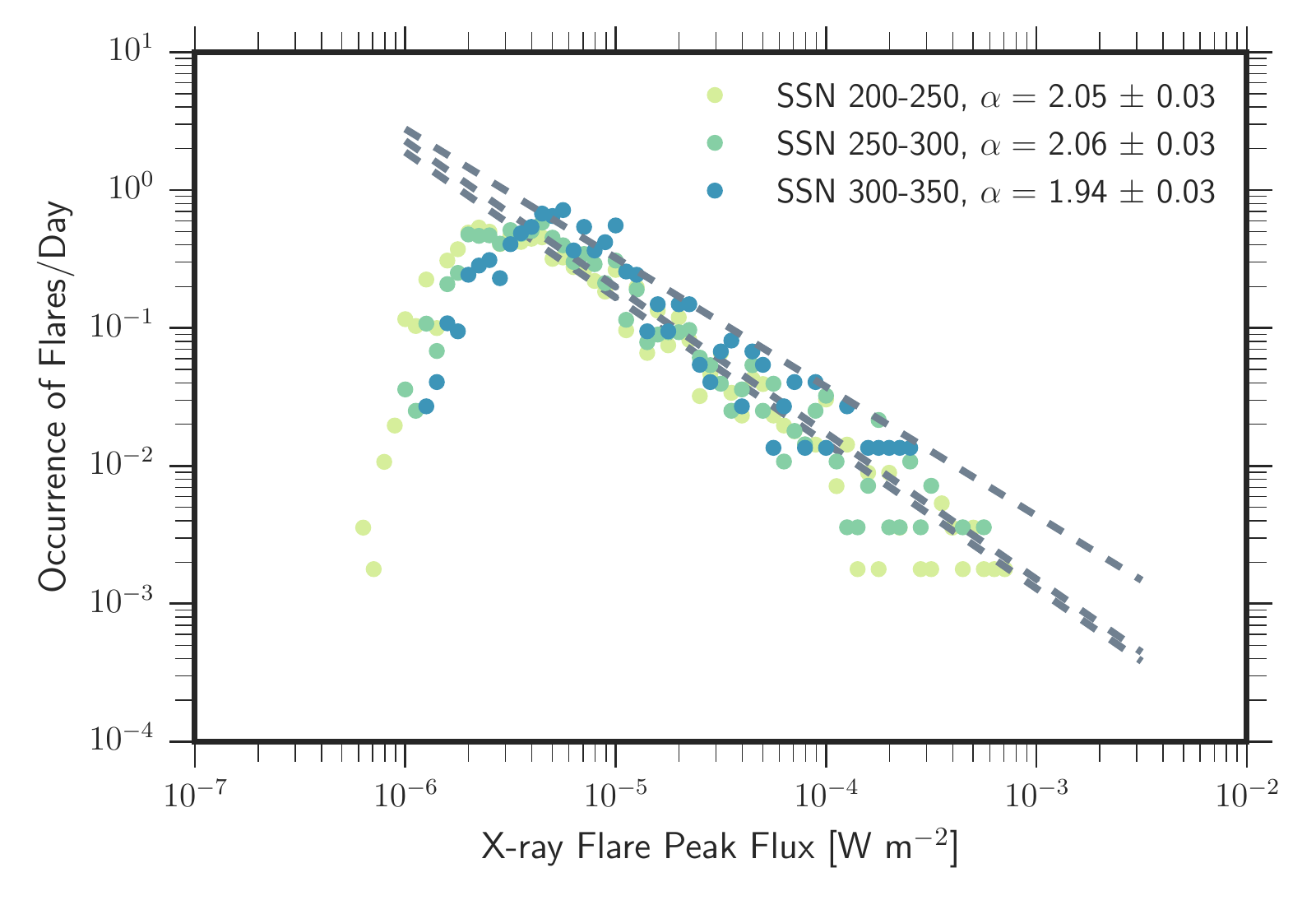}
\caption{X-ray flare rates as a function of sunspot number. The flare rates are computed for sunspot number bins from 0\,--\,350. When SSN is low, both the slope and normalization factors for the occurrence rate are low. The flare rates in the highest SSN ranges ($> 200$) are similar.}
\label{fig-flarerate}
\end{figure}

\indent\indent Flare rates and intensities are tied to solar activity levels. \citet{2011SoPh..274...99A} demonstrated this through their determination of the rates of the energetics, peak flux, and duration of X-ray flares in SC 21\,--\,23 using observations from GOES. Similarly, we determined the occurrence rate for flare peak flux during the rising phase to solar maximum and solar maximum periods using the NOAA flare list\footnote{The NOAA X-ray Flare Lists were downloaded from NOAA NGDC through the FTP site linked from here: \url{http://www.ngdc.noaa.gov/nndc/struts/results?t=102827\&s=25\&d=8,230,9}. } for SC 22\,--\,24 \citep{SWE:SWE20217}. Using the solar X-ray background analysis from \citet{2041-8205-793-2-L45}, the rise phases are defined as occurring from August 1986\,--\,August 1988 (SC 22), May 1996\,--\,May 1998 (SC 23), and December 2008\,--\,December 2011 (SC 24) and the solar maximum phases are defined as August 1988\,--\,August 1991 (SC 22), May 1999\,--\,May 2003 (SC 23), and December 2011\,--\,December 2014 (SC 24). We found that the occurrence rates of flares are consistent between the past three solar cycles. Figure~\ref{fig-flarerate} shows these results during the rise to solar maximum and solar maximum phases. We showed that the shape of the occurrence frequency distributions, which were fit as power-laws, differed between the phase of the solar cycle.

Expanding on the previous results, we present the X-ray flare occurrence frequency rates for a range of sunspot numbers. The SSN ranges are from 0\,--\,350 in bins of 50. For each X-ray flare, we use the daily sunspot number on the day of the flare. The results are presented in Figure~\ref{fig-flarerate}. The flare rates were fit with a power-law of the form \begin{math} {N}(F_{1-8 \rm \AA}) = N_1 (F_{1-8 \rm \AA})^{-\alpha+1} \end{math}, where {\it N} is the occurrence frequency (flares rate/day), $F_{1-8 \rm \AA}$ is the logarithm of flare peak flux [W\,m$^{-2}$], and the fit parameters include the normalization factor, $N_1$, and the power-law index, $\alpha$. We utilized the Levenberg\,--\,Marquardt least-squares minimization technique to determine the best-fit function parameters, fitting the occurrence rate where peak flux $> 10^{-6}$\,W\,m$^{-2}$ (SSN $< 200$) or $5 > 10^{-5}$\,W\,m$^{-2}$ (SSN $\ge 200$). We did not include the B-class flares (peak flux $< 10^{-6}$\,W\,m$^{-2}$) in these fits since the low end of the power-law distribution is near the GOES detection threshold. Goodness of fit is assessed with the $\chi^2$ statistic, defined as $\chi^2 = \Sigma({\rm data - model})^2/{\rm std}^2$, where the data are the frequency distribution values [$N$], the model is the power-law fit, and std is the standard deviation for each of the measurements of $N$. Good fits are those where $\chi^2$/dof is close to unity, where dof are the degrees of freedom or number of data points minus the number of free parameters that are fit.

Table~\ref{table-occurrence} includes the results of the analysis of X-ray flare occurrence rates. The normalization factor and slope of the occurrence rate clearly varies with SSN, as expected. The number of B-class or higher X-ray flares per day ranges from $< 2$ on the lowest SSN days to $\approx 9$ at the highest SSN. We find that the normalization factor scales with SSN, which is highest when SSN is highest. \citet{SWE:SWE20217} found that the slope is flatter at lower solar activity levels. In the highest SSN bins, the fits were less reliable (high reduced $\chi^2$). This { may explain} the flatter slope measured in the highest SSN bin. However, for SSN $> 200$, the slopes appear consistent in Figure~\ref{fig-flarerate}. As in \citet{SWE:SWE20217}, we expect additional differences in the normalization factors and slopes for the flare rates between the solar cycles. For instance, we had shown that the normalization varied by up to $\approx 10$\,\% and the slope varied by 9\,\% (solar maximum) to 16\,\% (rising phase) between SC 22, 23, and 24.

These flare rates allow for a comparison of expected historical rates with the revised SSN dating back several hundred years. Additionally, the SSN/X-ray background relationship derived in the previous section allows for calibration with similar solar-analogs. X-ray flux measurements of nearby solar stars, including $\alpha$ Cen  
({\it e.g.}, \citealt{2014AJ....147...59A}), unveil stellar magnetic activity cycles. Comparing the flux of these stars with the SSN relationship and flare rates provides new estimates of flare rates that can be used for instance in determining expected radiation effects for extra-solar planet habitability studies.


\begin{table}
\caption{Occurrence of X-ray Flares by SSN}\label{table-occurrence}
\begin{tabular}{llllll}
\hline
{\bf SSN range} & {\bf $N_{\rm Years}$} & {\bf $N_{\rm Flares}$} & {\bf $F_1$}  & {\bf $\alpha$} & {$\chi^2$/(dof)} \\
\hline
0-50 & 11.98 & 8252 & 	-6.94 $\pm$ 0.10 & 1.91 $\pm$ 0.06 & 26.2/39\\
50-100 & 5.92 & 12309 & -6.60 $\pm$ 0.04 & 2.01 $\pm$ 0.04 & 28.3/48\\
100-150 & 4.69 & 11803 & -6.69 $\pm$ 0.03 & 2.10 $\pm$ 0.03 & 23.6/47\\
150-200 & 3.24 & 8607 & 	-6.56 $\pm$ 0.02 & 2.11 $\pm$ 0.03 & 34.7/46\\
200-250 & 1.54 & 4525 & -6.05 $\pm$ 0.02 & 2.05 $\pm$ 0.03 & 83.6/45\\
250-300 & 0.76 & 2280 & -6.00 $\pm$ 0.02 & 2.06 $\pm$ 0.03 & 116.8/41\\
300-350 & 0.20 & 653 & -5.17 $\pm$ 0.02 & 1.94 $\pm$ 0.03 & 451.0/35\\
\hline
\end{tabular}
\vspace{0.25cm}

{\small
Occurrence rates were determined for SSN ranges. The number of years [$N_{\rm Years}$] and number of flares [$N_{\rm Flares}$] corresponding to the SSN ranges are given. Results from power-law fits to the occurrence distributions of the form $N(F_{1-8\,{\rm \AA}}) = N_1 \times (F_{1-8\,{\rm \AA}})^{-\alpha + 1}$ are included through the best-fit parameters and the $\chi^2$ statistic along with the number of degrees of freedom (dof).
}
\end{table}

\section{CME Rates}
\label{S-CME}     

\indent\indent Just as the properties of X-ray flares are tied to solar cycle, the rate
and properties of CMEs correlate with solar activity levels \citep{1994JGR....99.4201W,2009ApJ...691.1222R}. To determine dependence of the CME
properties on SSN, we used the SOHO LASCO Coordinated Data Analysis
Workshops (CDAW) CME catalog\footnote{A description of the LASCO CDAW catalog fields is available at \url{http://cdaw.gsfc.nasa.gov/CME_list/catalog_description.htm}} ({\it e.g.}, \citealt{2009EM&P..104..295G}). Alternatively, the automatically generated CACTUS CME list \citep{2009ApJ...691.1222R}, which is also based on SOHO LASCO data, could have been used. However, as there is no human supervision for CACTUS, the authors caution its use for statistical analyses. Therefore, we chose to utilize the CDAW catalog, which is created by the CDAW Data Center team. This introduces biases from changing criteria for classifying an event as a CME ({\it e.g.}, from changes in understanding of CMEs over the past 20 years), which we correct with a width filtering criteria that we explain in the following.  

\begin{figure}
\centering
\includegraphics[width=\linewidth]{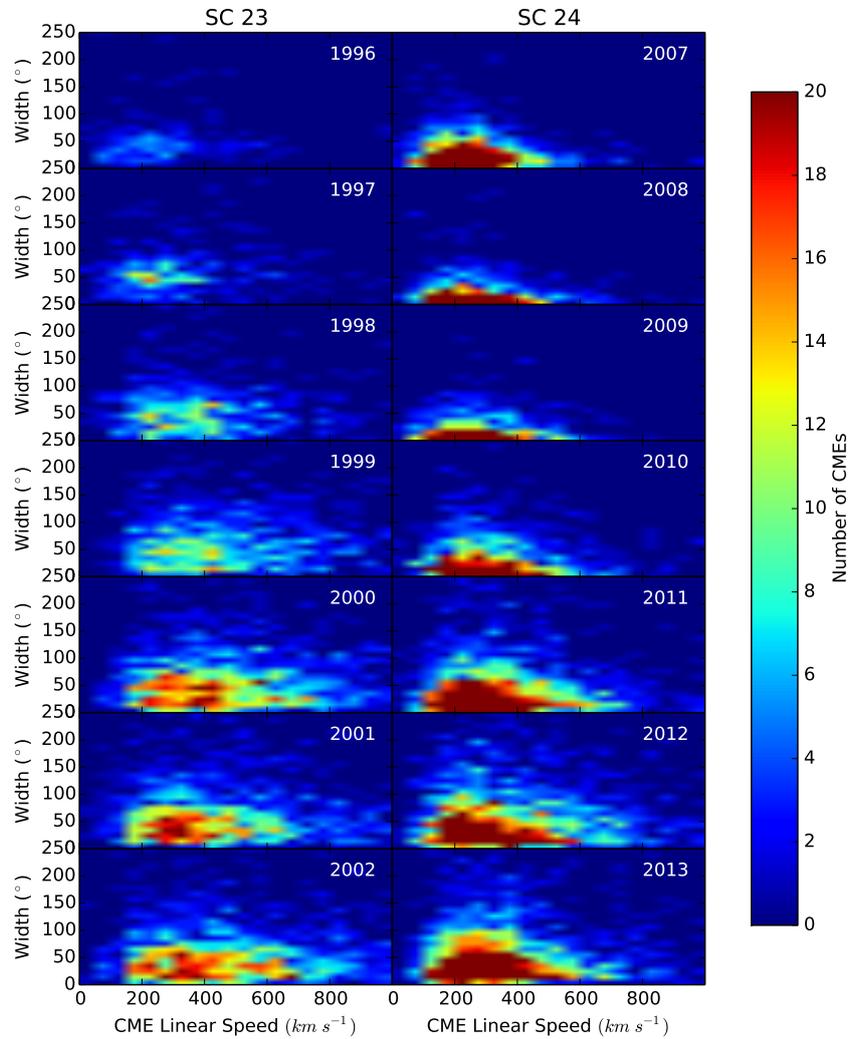}
\caption{\small Two-dimensional CME frequency histograms of width and speed from analysis of
the LASCO CDAW CME catalog. Solar minimum years
are at the top of the figure, leading down to solar maximum near the bottom, with a direct comparison of
similar phases of solar-cycle between Cycle 23 (left) and 24 (right). As the CME detection methods were improved, more CMEs are detected in SC 24 leading to higher number counts than in SC 23. }\label{fig-cme}
\end{figure}
\begin{figure}
\centering
\includegraphics[width=0.95\linewidth]{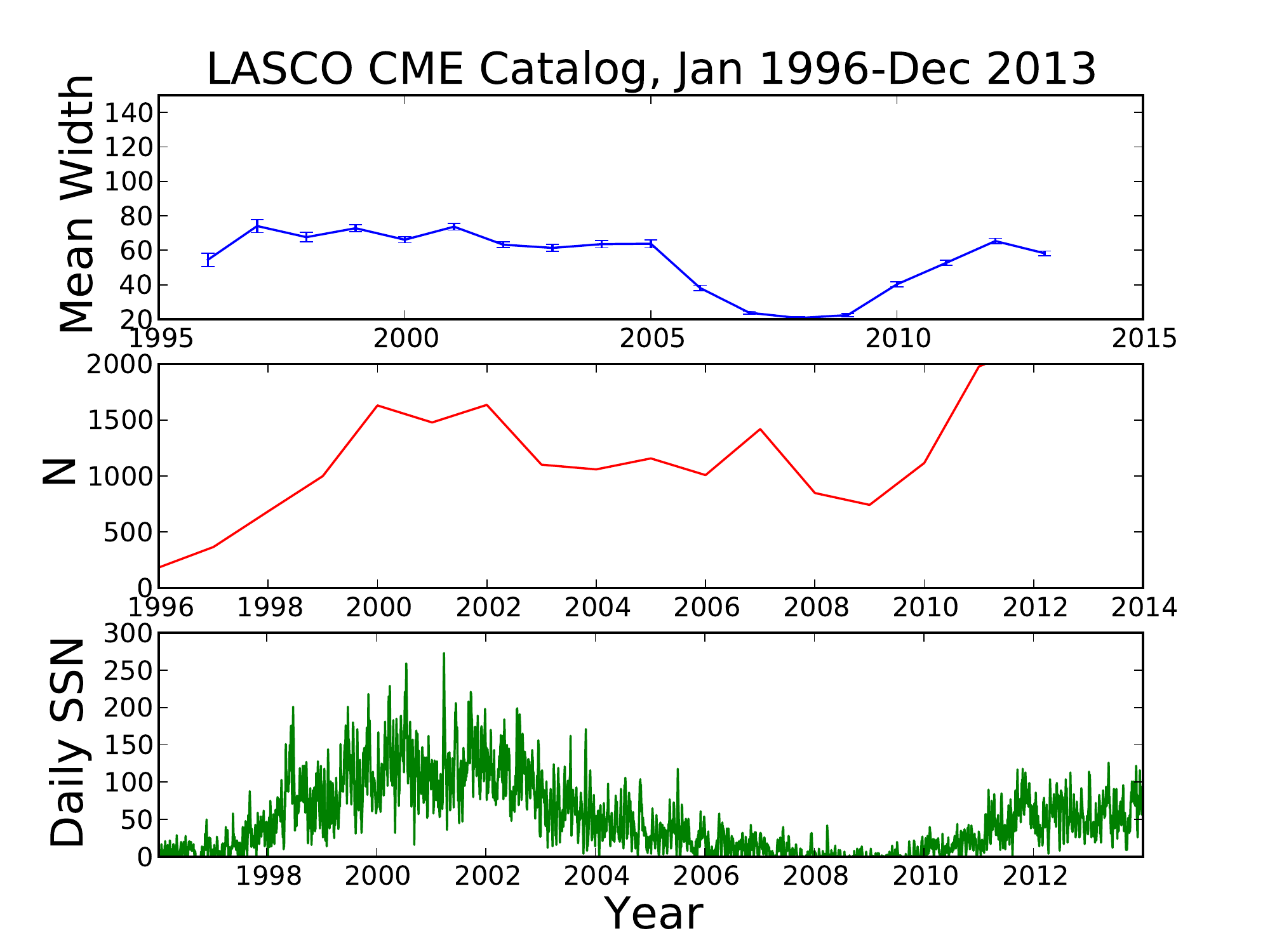}
\includegraphics[width=0.95\linewidth]{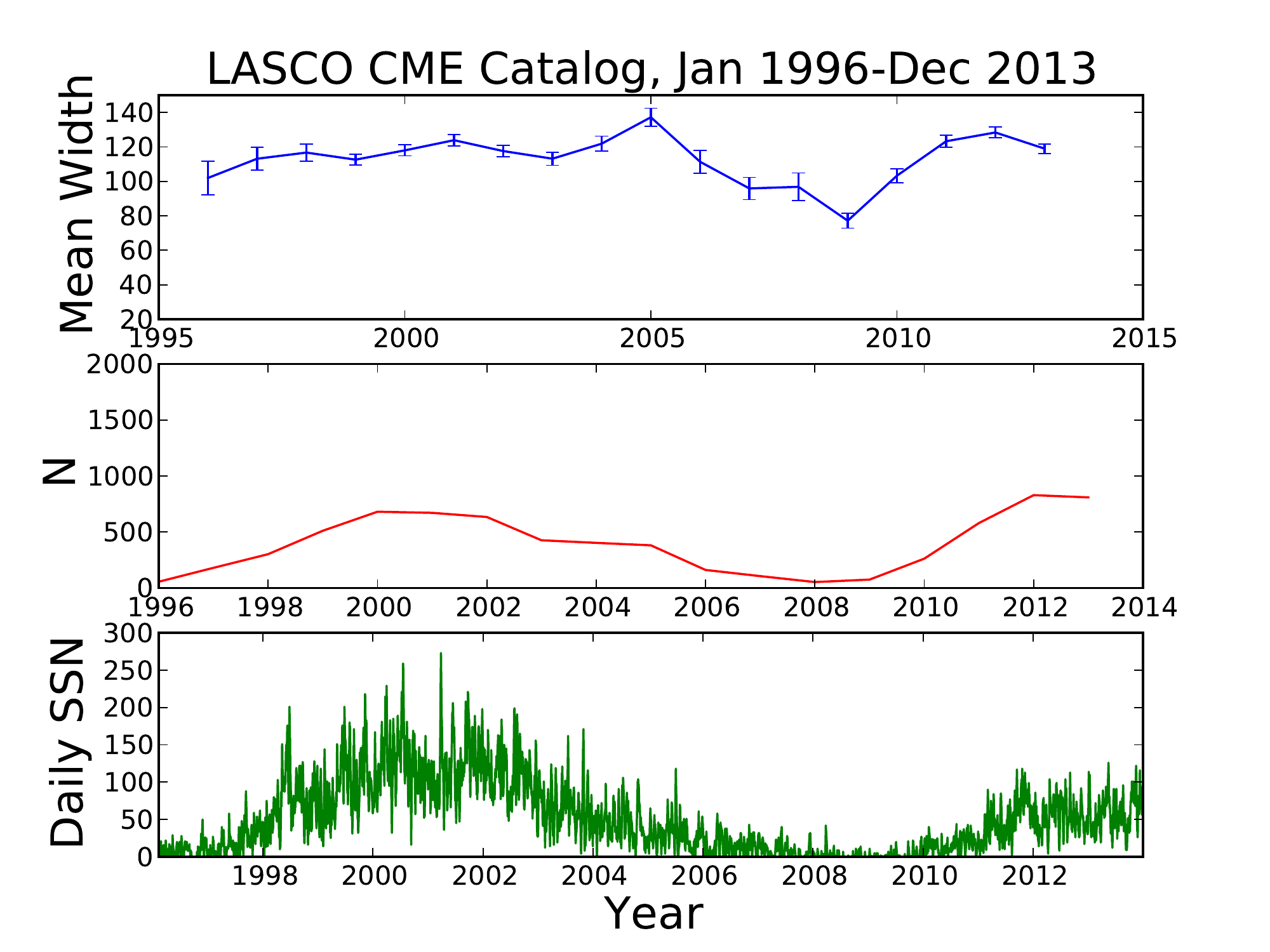}
\caption{\small CME width and frequency time series along with
daily-averaged SSN from the LASCO CDAW catalog. The plots show CME angular width before (top) and after (bottom) normalizing the catalog to include consistent CME selection criteria (excluding widths below 55$^{\circ}$) from SC 23 and 24. Consistent with earlier studies, we find that { in our filtered sample,} the number of CMEs increases toward solar maximum along with the angular width.}\label{fig-dailyssncmewidth}
\end{figure}
\begin{figure}
\centering
\includegraphics[width=0.95\linewidth]{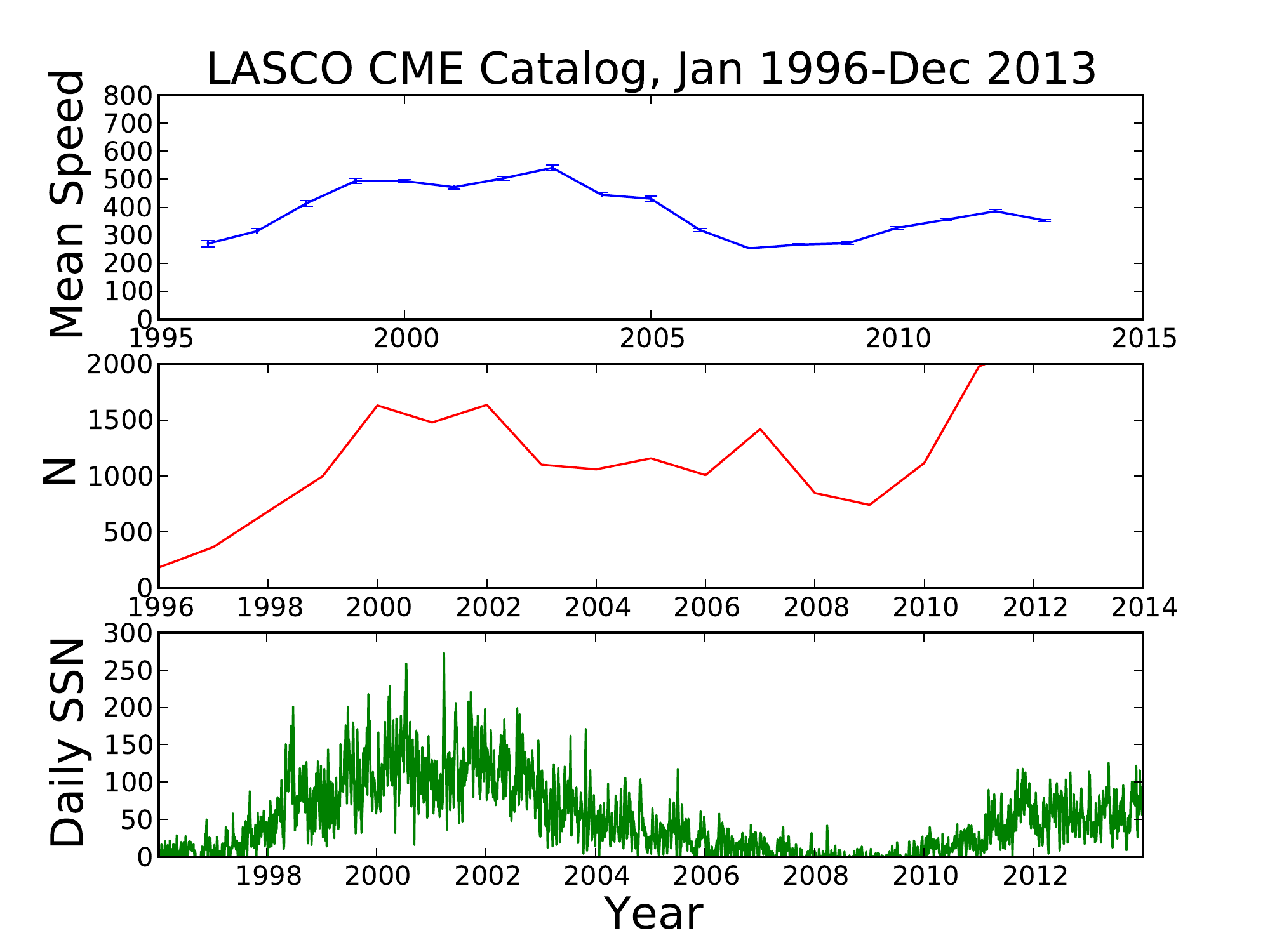}
\includegraphics[width=0.95\linewidth]{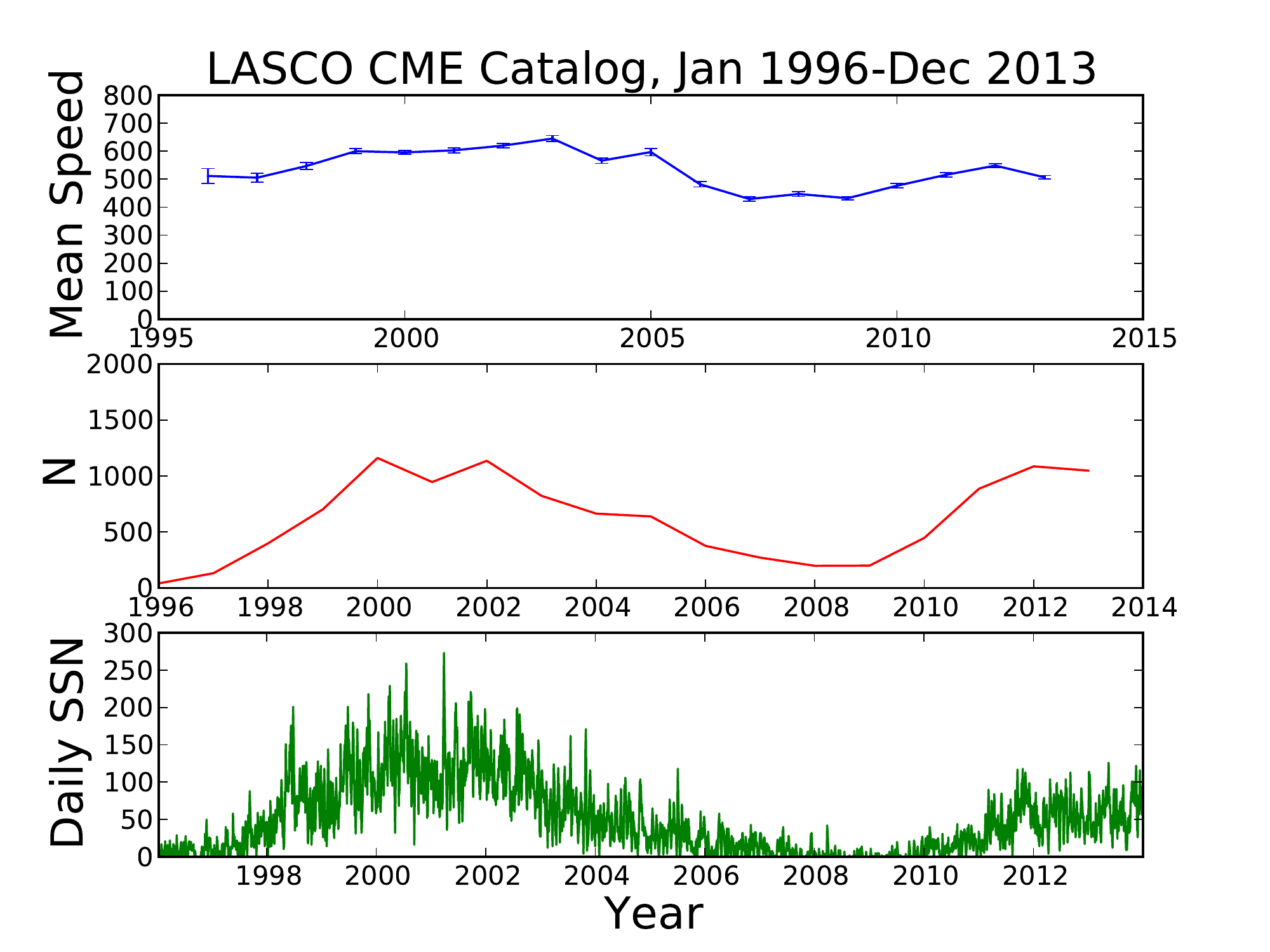}
\caption{\small CME speed and frequency time series along with
daily-averaged SSN to illustrate solar cycle dependence. CME projected linear speed is shown before (top) and after (bottom) normalizing the CDAW catalog (as in Figure~\ref{fig-dailyssncmewidth}). Average speed increase toward solar maximum in our filtered sample.}\label{fig-dailyssncme}
\end{figure}

Based on previous studies \citep{2000JGR...10518169S, 2004JGRA..109.7105Y}, the frequency of CMEs varies with solar cycle ({i.e.}, direct correspondence between solar cycle phase and number of CMEs).  
However, the LASCO catalog does not initially exhibit this behavior. This is due to inconsistency in the CMEs that are included, such that more weaker CMEs are included in the later years of the catalog. Figures~\ref{fig-cme} --~\ref{fig-dailyssncme} illustrate this effect, where the number of detected CMEs is higher in
SC 24 (right) than SC 23 (left). { The parameters shown include the LASCO catalog's {\it linear speed} and {\it angular width}. The LASCO linear speed measurements result from fitting a line through the CME height-time data (i.e., linear fits to coronagraph plane of sky speeds). The LASCO width measurements are sky-plane widths (i.e., position angle extent of CMEs which are upper limits on the true CME width). } The LASCO data shows that the mean CME width drops significantly starting just prior to SC 24. To correct for the effect of differences in catalog criteria over the years, we introduced a width filter to homogenize
the LASCO catalog through analysis of the statistical properties of the
more than 20,000 CMEs from 1996\,--\,present.  

{ We found that using a width criterion of 55$^{\circ}$ resulted in consistent number counts of CMEs between similar phases of SC 23 and 24. We determined the width criterion by computing the mean width of CMEs in the LASCO catalog for each year, as shown in Figure~\ref{fig-dailyssncmewidth}. Without the filtering, this figure shows that no trends are seen between the number of CMEs and the sunspot cycle. We computed the mean width by fitting a linear function to the yearly mean width, which yields a mean width of 55$^{\circ}$. Once this criterion is applied as a filter to the LASCO catalog, the expectation of the number density of CMEs varying with solar cycle ({\it e.g.}, 
\citealt{2000JGR...10518169S, 2004JGRA..109.7105Y}) is observed as shown in Figures~\ref{fig-dailyssncmewidth} and \ref{fig-dailyssncme} (bottom). Our filtered LASCO sample size includes 7151 CMEs.}

Figure~\ref{fig-cme} shows the solar-cycle dependence of both CME width and
linear speed that \citet{2004JGRA..109.7105Y} and \citet{2009Ap&SS.323..135M} found.  
Progressing through either solar cycle (from solar minimum near the top of Figure~\ref{fig-cme} to solar maximum at the bottom), the density of CMEs shown through
the contours expand along both the width and speed axes, suggesting that
CMEs on average are larger and faster during solar maximum years
relative to minimum years.  As shown in Figures~\ref{fig-dailyssncmewidth} and~\ref{fig-dailyssncme}, this
variation correlates with daily-averaged SSN.  CME
frequency as a function of year after catalog normalization is also plotted. We find that linear speed, width,
and CME frequency exhibit a significant direct correlation with SSN. { Interestingly, even though SC 24 has a much lower SSN during maximum, we find that the number of CMEs is roughly consistent with or higher than the number of solar cycle 23 CMEs during similar phases in the cycles. \citet{2015ApJ...804L..23G} recently reported similar results when comparing halo CMEs between SC 23 and 24.}


\begin{table}
\caption{CME Statistics from Solar Cycle Rise to Maximum}\label{table-cmestats}
\begin{tabular}{llllll}
\hline
{\bf SSN range} & {\bf $N_{\rm CMEs}$} & {\bf Linear Speed [km s$^{-1}$]} & {\bf Width [$^{\circ}$]}  \\
\hline
Solar Cycle 23 & Jan. 1996 & Mar. 2001 \\
\hline
0-50 & 245 & 366.97 $\pm$ 258.10 & 111.84 $\pm$ 85.27\\
50-100 & 407 & 485.08 $\pm$ 330.74 & 114.65 $\pm$ 83.58\\
100-150 & 433 & 490.07 $\pm$ 313.62 & 116.81 $\pm$ 79.70 \\
150-200 & 422 & 506.74 $\pm$ 270.78 & 114.18 $\pm$ 76.52 \\
200-250 & 260 & 528.12 $\pm$ 311.14 & 122.91 $\pm$ 85.40\\
250-300 & 70 & 535.13 $\pm$ 281.28 & 123.36 $\pm$ 82.69\\
300-350 & 7 & 556.71 $\pm$ 356.15 & 108.57 $\pm$ 57.42\\
\hline
Solar Cycle 24 & Jan. 2008 & Dec. 2013 \\
\hline
0-50 & 654 & 360.88 $\pm$ 257.12 & 107.51 $\pm$ 72.09\\
50-100 & 881 & 436.15 $\pm$ 274.19 & 124.91 $\pm$ 84.46\\
100-150 & 901 & 446.05 $\pm$ 293.18 & 122.89 $\pm$ 82.97\\
150-200 & 173 & 468.39 $\pm$ 329.80 & 122.73 $\pm$ 82.42\\
\hline
\end{tabular}
\vspace{0.25cm}

{\small
Statistics for the solar cycle phases from rise to maximum for both SC 23 and 24 by SSN. The number of CMEs on days within the indicated SSN range ($N_{\rm CMEs}$), as well as average and standard deviations of the linear speed and width. All CMEs with widths below 55$^{\circ}$ are excluded to result in a more uniform comparison (i.e., more narrow CMEs are included in later versions of the CDAW LASCO CME catalog).
}
\end{table}

In Table~\ref{table-cmestats}, we present statistics on a comparison of the number, speed, and width of CMEs from SC 23 and 24. Since SC 24 is not complete, we compare similar phases of the solar cycles from the rise from minimum to maximum. As in Section~\ref{S-flares}, the daily SSN matching each CME is used to find the statistics within a range of SSN. From our comparison of the two cycles, we find that the filtering to normalize the CME catalog may need adjustment, since there are 1844 CMEs recorded in SC 23 and 2609 in SC 24. At lower SSN ($< 150$), there are significantly more CMEs recorded in SC 24. Many of these additional CMEs have slower speeds than the SC 23 CMEs, though the average width tends to be the same or larger in SC 24. As the SSN increases, there are significantly more CMEs in SC 23 (i.e., from SSNs of 150\,--\,200 there are 2.4 times more CMEs in SC 23). The speed is also higher in SC 23, but the average widths are consistent. When we compare the average speeds and widths from different SSN ranges, we find that speed increases with SSN, while average width is fairly consistent. 

One caveat on this analysis is that we compare CME properties with no lag between solar activity and CMEs.   
\citet{2004ASSL..317..201G}, for instance, found that the peak in the CME rate lagged the peak in sunspot number by months. Additionally, \citet{2011ApJ...727...44K} found that CME speed lags SSN by several months. We have not investigated the presence of a time lag between CME properties and SSN in the current analysis, but note that other studies do see such effects. We defer a more detailed analysis of time lag affects to future studies. 

\begin{figure}
\includegraphics[width=0.9\linewidth]{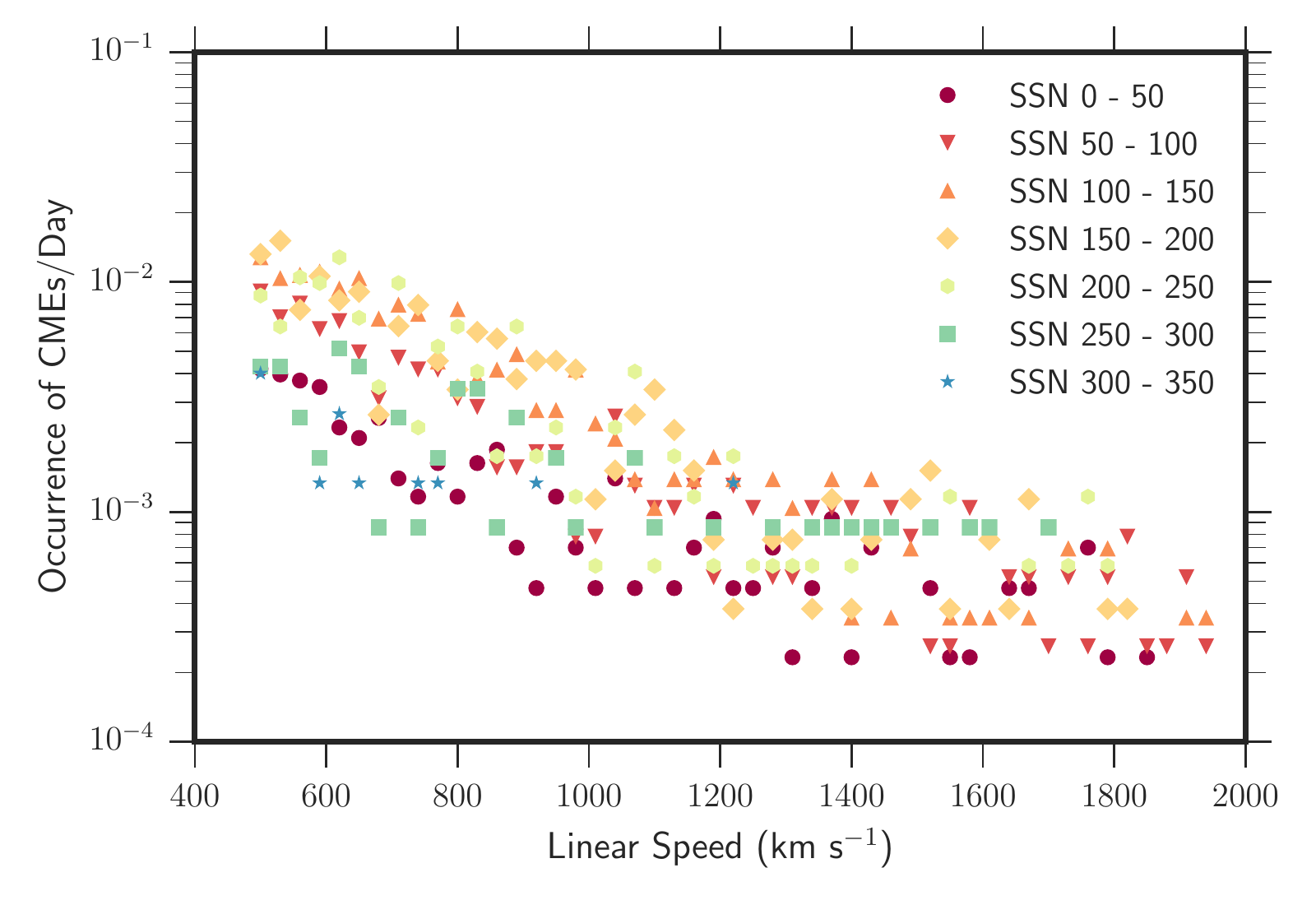}
\includegraphics[width=0.9\linewidth]{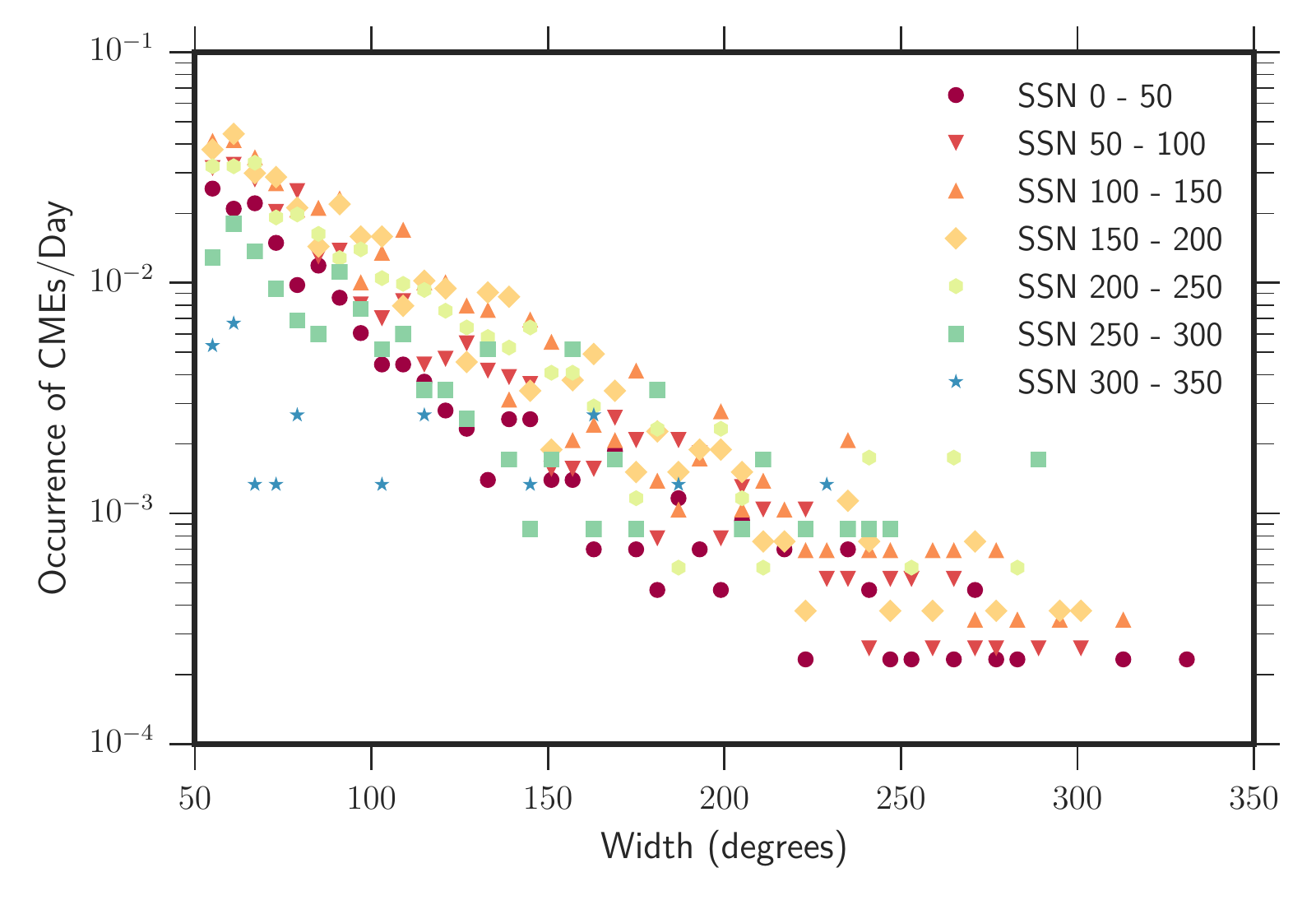}
\caption{\small Frequency distributions for CME linear speed and width using SC 23 observations. The normalization factor varies between SSN distributions, while the slope remains constant, for both distributions.
}\label{fig-cmedistributions}
\end{figure}

As an additional investigation, which is not dependent on SSN, we compute the distribution of CME speed and width.  We use the data from SC 23, which spans a full solar cycle. Figure~\ref{fig-cmedistributions} presents the results. A power-law fit to the linear speed distribution yields: \begin{equation} N = (1.11 \pm 1.04) \times {\rm Linear\,Speed\,[km\,s^{-1}}]^{-0.015 \pm 0.005}, \end{equation} with correlation coefficient $R^2$ of 0.75 and p-value $\ll 10^{-10}$. This relation holds where SSN $<$ 250, since few measurements exist for the highest SSNs. Meanwhile, a power-law fit to the width distribution results in: \begin{equation}N = (1.21 \pm 1.04) \times {\rm Width\,[^\circ}]^{-0.035 \pm 0.007},\end{equation} with $R^2$ of 0.75 and p-value $\ll 10^{-10}$.


\begin{figure}
\includegraphics[width=0.9\linewidth]{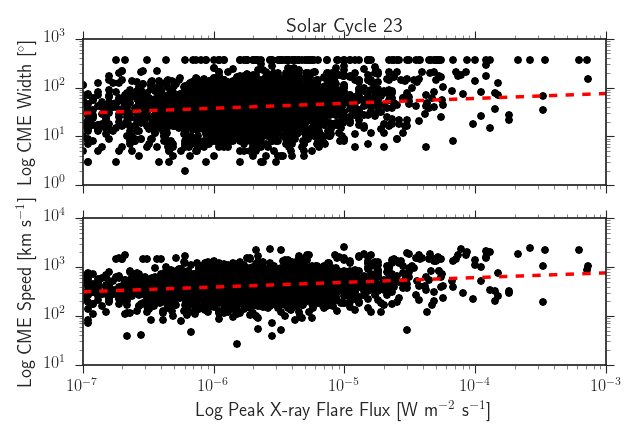}
\includegraphics[width=0.9\linewidth]{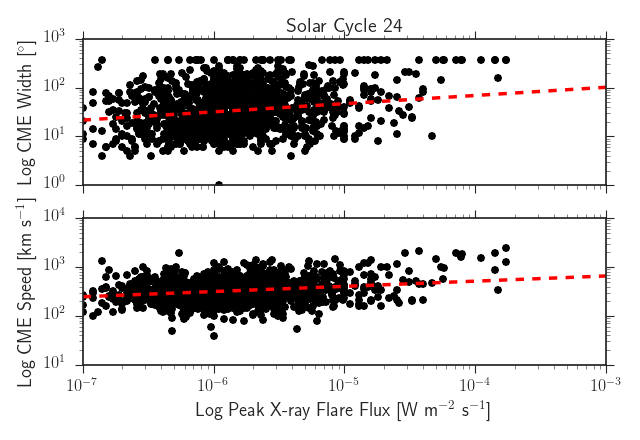}
\caption{\small Comparison of the CME width and speed to the X-ray flare peak flux for CMEs occurring within one hour following an X-ray flare. Both CME linear speed and width increase with increase X-ray peak flux.  The best-fit linear regression for each of these comparisons (shown as a dashed red line) are not statistically significant, with $R^2 \approx 0.05$.
}\label{fig-xraycme}
\end{figure}

Finally, we compare the CME speed and width to X-ray peak flux for all CMEs within one hour of the onset of an X-ray flare. Figure~\ref{fig-xraycme} shows the results for both solar cycles 23 and 24. We find that the highest flux X-ray flares are also associated with faster and wider CMEs. This is consistent with earlier results ({\it e.g.}, SC 23 CME analysis by \citealt{2002ApJ...581..694M}). We note that while a trend of increasing width and speed is found, there is no significant statistical relationship from a linear regression analysis ($R^2 < 0.1$).
   
\section{Summary} 
      \label{S-Conclusion} 
{      We have presented an analysis of revised sunspot number and have compared it to the fluctuation of the X-ray background.  While the X-ray background during the last solar minimum was often below the GOES detection limit, we show that a linear relationship exists between sunspot number and X-ray background flux.  This relationship  exhibits a slow steepening of the number spectrum moving from less active to more active solar cycles.  We also present the power-law relationships between X-ray flare rates binned by sunspot number ranges, showing that the flare rates are somewhat similar irrespective of the SSN.

A two-dimensional plot of the CME frequency shows that number counts of CMEs and uniform selection criteria are important in distinguishing the changes in the number and properties of CMEs from cycle to cycle. From a selected sample of CMEs, using a position angle measure of width $> 55^{\circ}$, we show that the number of CMEs in SC 24 is consistent with the number of CMEs during SC 23 at the same phase.  This result agrees with an analysis of halo CMEs from \citet{2015ApJ...804L..23G}. Additionally, we determine the power-law relationship between the projected CME speed and width as a function of sunspot number, showing that these parameters exhibit similar frequencies throughout a solar cycle.

In summary, our results include the relationship between SSN and X-ray background flux, power-law relationships of flare occurrence rate as a function of SSN, CME number counts through SC 23 and 24, and CME rates as a function of SSN. With the newly revised SSN dating back several hundred years, our results allow for a comparison of the expected flare and CME properties during historical times where X-ray and CME observations are not available. Our results also allow for a comparison of the X-ray background level where only SSN is estimated. This can provide insight for modeling of the solar conditions associated with extreme historical storms, like the 1859 Carrington event. Additionally, our results can be used to estimate CME and flare rates in solar-analogs, to estimate stellar activity levels from X-ray flux measurements and from this radiation effects on extra-solar planets.
}

\acknowledgements{
K.S. Balasubramaniam was supported by the Air Force Office of Scientific Research for research on "Physics of Coupled Flares and CMEs." L.M. Winter and R.L. Pernak were supported, in part, by AFRL contract FA9453-11-C-0231. The CME catalog used in this study was generated and maintained at the CDAW Data Center by NASA and The Catholic University of America in cooperation with the Naval Research Laboratory.
}

\section*{Disclosure of Potential Conflicts of Interest}
The authors declare that they have no conflicts of interest.

\bibliographystyle{spr-mp-sola} 

\end{article} 

\end{document}